\documentclass[a4paper,unsortedaddress,twocolumn,superscriptaddress,
groupedaddress,aps,prl,reprint]{revtex4-1}

\usepackage{graphicx}
\usepackage[T1]{fontenc}
\usepackage{lmodern}
\usepackage{braket}

\usepackage{amsmath,amsfonts,amssymb}
\newcommand{\dd}{{\rm d}}

\newcommand{\ii}{\mathrm{i}}

\newcommand{\be}{\begin{equation}}
\newcommand{\ee}{\end{equation}}
\newcommand{\ba}{\begin{eqnarray}}
\newcommand{\ea}{\end{eqnarray}}

\newcommand{\met}{SESM}

\begin{document}
%\graphicspath{{figures/}}

\title{Identification of Resonant States via the Generalized Virial Theorem}

\author{Luigi \surname{Genovese}}
\email{luigi.genovese@cea.fr}
\affiliation{Univ. Grenoble Alpes, INAC-SP2M, F-38000 Grenoble, France //
        CEA, INAC-SP2M, Laboratoire de simulation atomistique (L\_Sim), Grenoble, F-38000, France}

\author{Alessandro \surname{Cerioni}}
\affiliation{Univ. Grenoble Alpes, INAC-SP2M, F-38000 Grenoble, France //
        CEA, INAC-SP2M, Laboratoire de simulation atomistique (L\_Sim), Grenoble, F-38000, France}

\author{Maxime \surname{Morini\` ere}}
\affiliation{Univ. Grenoble Alpes, INAC-SP2M, F-38000 Grenoble, France //
        CEA, INAC-SP2M, Laboratoire de simulation atomistique (L\_Sim), Grenoble, F-38000, France}

\author{Thierry \surname{Deutsch}}
\affiliation{Univ. Grenoble Alpes, INAC-SP2M, F-38000 Grenoble, France //
        CEA, INAC-SP2M, Laboratoire de simulation atomistique (L\_Sim), Grenoble, F-38000, France}

\date{\today}

\begin{abstract}
The numerical extraction of resonant states of open quantum systems is usually a difficult problem.
Regularization techniques, such as the mapping to complex coordinates or the addition of Complex Absorbing Potentials 
are typically employed, as they render resonant wavefunctions localized
and therefore normalizable. 
Physically relevant metastable states have 
energies that do not depend on the chosen regularization method.
Their identification therefore involves cumbersome comparisons between
multiple regularised calculations, often performed graphically, which require fine-tuning and specific intuition 
to avoid approximated, if not wrong, results.
In this Letter, we define an operator that explicitly measures such invariance, valid for any arbitrary mapping of 
spatial coordinates. Resonant states of the system can eventually simply be identified evaluating the expectation value 
of this operator. 
Our method 
eases the extraction of resonant states even for 
numerical potentials that are difficult to scale to complex coordinates, and avoids the 
need for \textit{ad hoc} complex absorbing potentials.
We provide explicit evidence of our findings discussing one-dimensional case-studies,
also in the presence of external electric fields.
\end{abstract}

\maketitle

%The concept of resonance is ubiquitous in all areas of Quantum Physics.
Resonant states % (also called ``Gamow vectors'' or ``Siegert states''), 
are ubiquitous in Quantum Physics. 
Also referred to as ``Gamow vectors'' or ``Siegert states'',  %and 
they can be defined as solutions of the time-independent Schr\"odinger equation 
subject to outgoing boundary conditions.
Described at first by Gamow \cite{Gamow} via quasi-stationary states, the concept of resonant states has been
widely developed in the field of atomic and nuclear physics
(see e.g.\ Ref.~\cite{PhysRevC.47.768}), then adopted for the analysis of scattering properties of quantum systems with open boundaries~\cite{Hatano2008}.
%and, more generally, for all quantum systems .
Various literature has shown that Siegert states \emph{encode}
in compact form the response properties of a system \cite{Myo01051998,Lind1994}.
In particular, the analytic structure of the resolvent operator (i.e. the Green's function) is completely 
determined by resonant energies and wavefunctions. % (also called ``Gamow vectors'' or ``Siegert states'').
In the words of Ref.~\cite{PhysRevLett.79.2026}, resonant states expansions offer the ``possibility of a unified
description of bound states, resonances, and continuum spectrum in terms of a purely discrete set of states''.
For one-body Hamiltonians of quantum systems, the identification of their resonant energies and wavepackets
is therefore of paramount importance.

%However, b
%Being eigenstates
%Due to their complex energy, 
Resonant wavefunctions exhibit complex wavenumbers
$k = k_R - \ii\, k_I$ and divergent asymptotes $\propto \exp({\ii\,k \cdot x})$.
As such, they are  not to be found in the Hilbert space of square-integrable functions.
Yet, they find rigorous mathematical foundations in the domain of non-Hermitian Quantum Mechanics~\cite{moiseyev2011non}.
In order to ease their numerical treatment, methods 
have been proposed, which render resonant states square-integrable, allowing their 
computation under bound-state-like ({\it i.e.}\ Dirichlet) boundary conditions. 
These regularization methods imply a \emph{modification} of the original Hamiltonian of the system,
driven by a set of continuous parameters (usually called $\eta$ or $\theta$, see e.g.~\cite{HeadGordon}),
which eventually leads to %the objective of making the operator 
a complex-valued operator, {\it i.e.}\ \emph{explicitly} non-Hermitian.
In this way, localized -- thus square-integrable -- eigenvectors with complex eigenvalues may show up in the spectrum of the resulting operator. %exist. 
In this context, \emph{physically relevant} resonant states are those having an energy that is \emph{invariant}
with respect to the chosen regularization method.

The identification of resonant energies requires comparisons between
multiple regularised calculations, typically performed graphically (via so-called $\eta$- or $\theta$-trajectories).
Such a comparison is generally difficult to be performed and requires highly precise calculations 
to identify stable points in the spectra of several complex Hamiltonians. 
This is especially true when the potential is only known numerically.
Moreover, even when a stable point is found, little is known about the numerical quality of the 
corresponding eigenvector, that may depend on computational parameters such as the size of simulation box or
the choice of the numerical basis set.

In this Letter, we revisit the properties of Siegert states under \emph{arbitrary} parametric transformations
of spatial coordinates. %, %parametrized by a multi-index $\xi$, 
%by specifically focusing on their energy. 
We eventually introduce an operator whose quantum expectation value is explicitly associated to the variation
of the energy with respect to the parameters of the coordinate mapping. %$\xi$. 
%We relate this 
Such operator can be related to a generalisation of the classical Virial Theorem for stationary states.
This procedure provides a reliable and rigorous approach to identify resonant states
without the need neither of explicit variations of the parameters nor
the analytic continuation of the numerical potential in the complex plane.
%Being derived on analytical basis, this method makes the usage of (exterior) coordinate scalings
%much more powerful than the adding of Complex Absorbing Potentials.
%Moreover, 
%it provides a natural indication of ``quality'' of states with complex energies, 
The method we propose  is crucial
%useful 
in indicating which states have some
physical interest and, at the same time, provides an estimate of the accuracy of the computational treatment.
By presenting some illustrative examples involving one-dimensional models, we demonstrate that our approach %this information 
can be used
to single out the states that are more relevant in determining the linear response of  open quantum systems.

In order to motivate the interest of our results, %findings, 
let us first illustrate the main advantages and drawbacks of the most popular regularization techniques.
In a number of numerical investigations, resonant states are computed through the 
Complex Absorbing Potential (CAP) method \cite{Rom1991199}, where
a complex potential is added to the Hamiltonian such as to \emph{absorb} the 
decaying particle described by the outgoing resonant state. 
Within this approach, the original one-body potential is not modified, which 
makes this approach suitable for numerical potentials~\cite{SourceSink}.
A resonant state is then identified as a CAP-independent state,
and its energy is often found by verifying numerically its invariance with respect to variations of the CAP strength (the $\eta$-trajectories).
However, there is no unambiguous recipe for the CAP to ensure that some resonant states
appear in the spectrum of the non-Hermitian Hamiltonians. 
The absorbing boundary described by the CAP may induce artificial ``reflections'' of the resonant state wavefunction 
at the boundaries of the simulation domain, thereby altering their energy as well as their shape within the quantum device.
%Within the CAP method, t

A method which is based on rigorous mathematical foundations is the well-known
Complex Scaling Method (CSM)~\cite{Aguilar1971}, in which spatial 
coordinates are ``scaled'' by a complex factor, $x \rightarrow x e^{\ii\, \theta}$. 
All the resonant states for which $k_I / k_R  < \tan\theta$ become localized, and their 
%The so-called Complex Virial theorem guarantees that s
%Such resonant states 
energies are $\theta$-independent.
%whereas c
Instead, continuum  states show up along straight lines, 
rotated by an angle $-2\theta$ with respect to the real axis. 
If the Hamiltonian has only one threshold energy ({\it i.e.} $V(\infty)$, see Ref.~\cite{CGDD2013}), 
resonant energies can be identified by simply looking at their position 
with respect to the rotated continuum. However, %very often also 
even for the CSM, the identification of a resonant energy
often relies on its numerical invariance with respect to $\theta$ ($\theta$-trajectories).

Despite its conceptual simplicity, CSM is \emph{unfit} for 
the treatment of generic numerical potentials with large spatial extension, as it introduces 
high-frequency oscillations of the potential far from the fixed point of the scaling transformation.
This is true \emph{even for analytic potentials}.
To illustrate this point, it is enough to consider the simplest prototype of a localized, smooth function, 
a Gaussian $G(\sigma,x_c;x)$, %and discuss how it changes 
undergoing a complex scaling transformation centered at the origin:
\be\label{gaussian}
G(\sigma,x_c;x e^{\ii\theta}) = \exp\left[- \frac{(x e^{\ii \theta}-x_c)^2}{2\sigma^2} \right]\;.
\ee
Such function can model a ``diffusive center'' placed at the position $x_c$ in
the simulation domain.
When $x_c/\sigma \gtrsim 1$, as in spatially extended systems like electronic potentials with several diffusive centers,
the complex scaling transformation induces high-frequency (as well as high-amplitude) oscillations on the potential, 
very difficult to be captured numerically.
The potential becomes so oscillating that an accurate numerical treatment is unfeasible even for computational domains of moderate size: 
the numerical basis set should be able to capture the large and rapid oscillations \emph{both} of the potential \emph{and} of
the eigenvectors, making the computational cost overwhelming.

An elegant generalization of the CSM exists, still based on rigorous foundations.
Referred to as ``Reflection-Free CAP'' in Ref.~\cite{0953-4075-31-7-009}, 
%in the seminal paper by Moiseyev 
%this approach is also known as  
or as ``Smooth Exterior complex Scaling Method'' (\met), this approach
somehow couples the CAP method with the CSM.	
The \met\  stems from the coordinate transformation 
\be\label{eq:GenCS}
x\rightarrow F_\xi(x)\;,
\ee
$\xi=0$ being the identity transformation.
A rigorous, non-Hermitian Hamiltonian can be obtained out of the reparametrization $F_\xi(x)$, from which resonant states can be extracted.
The function $F_\xi(x)$ is generally chosen to
%which depends on a number of parameters, 
tend asymptotically to $x\, e^{\ii \theta}$ when $x\rightarrow\pm\infty$, thereby 
reconciling  with the CSM. 
When the family of functions can be chosen so that $F_\xi(x)=x$ where $V(x)\neq 0$,
the potential of the \met\ Hamiltonian can be left unscaled.
%This fact leads to a couple of important advantages: 
%1.\ A rigorous, non-Hermitian hamiltonian can be defined from the 
%function $F_\xi(x)$, from which the resonant states can be extracted;
%2.\  Original potentials can be used \emph{as-is}, without 
%special treatments.
%3.\ 
Also non-local ({\it e.g.}\ many-body) potentials, whose analytic continuation to complex coordinates might be cumbersome, 
can be treated with this method.
As in the CSM, resonant state energies have to be $\xi$-independent.
However, as in the CAP method, continuum states cannot be easlily excluded, %within the\met\ approach, 
hence resonant energies have to be found by explicitly verifying their independence 
with respect to the $\xi$ parameter space (typically, only $\theta$ is considered~\cite{:/content/aip/journal/jcp/127/3/10.1063/1.2753485}). 
%to simplify the treatment).

For multi-centered potentials, methods like CAP or \met\ seem very interesting, as
spatial coordinates can be left unscaled in the inner region of the simulation domain, whose extension is related to that of the potential.
%hat can be chosen at will.
In particular, the bound states whose support is contained in the unscaled region are not modified.
%This consideration is particularly interesting
This fact has a remarkable practical consequence: bound states of Hamiltonian can be first extracted with $\xi=0$,
then regarded as \emph{exact} eigenstates of the \met\ Hamiltonian, provided that their support is within 
the unscaled region.

Aside from the SESM or CSM, let us now consider the coordinate mapping of Eq.~\eqref{eq:GenCS} on a 
completely general ground, by assuming a generic form of the function $F_\xi$.
In what follows we adopt the notation
\be
\dot{F_\xi}(x) \equiv \frac{\partial F_\xi(x)}{\partial \xi}, \qquad f_\xi(x) = F_\xi'(x) \equiv 
\frac{\partial F_\xi(x)}{\partial x}\, .
\ee

For the normalization of bound states to be preserved, wavefunctions have to transform as 
$\psi \rightarrow \psi_\xi=\sqrt{f_\xi} \psi(F_\xi)$. %, where $ f_\xi(x) = F'_\xi(x)$.
Such transformation induces the following
modification on the position  and  %\emph{kinetic} term (cf.\ Ref.\ \cite{0953-4075-31-7-009}):
momentum operators, $\hat X$ and $\hat P = -\ii \overrightarrow{\partial_x} = \ii \overleftarrow{\partial_x}$  (cf.\ Ref.\ \cite{0953-4075-31-7-009}):
\begin{align}\label{eq:H_SES(M)}
\hat X &\rightarrow \hat X_\xi = \hat X[F_\xi] = F_\xi(\hat X)\;,\\
\hat P &\rightarrow \hat P_\xi =  \hat P[f_\xi] = -\ii 
\frac{1}{\sqrt{f_\xi}}  \overrightarrow{\partial_x} \frac{1}{\sqrt{f_\xi}} = \ii 
\frac{1}{\sqrt{f_\xi}}  \overleftarrow{\partial_x} \frac{1}{\sqrt{f_\xi}}\;. \notag
%-\frac{1}{2} \frac{1}{\sqrt{f_\xi}} \partial_x \frac{1}{f_\xi} \partial_x \frac{1}{\sqrt{f_\xi}} 
%=\\=-\frac{1}{2}\nabla^2 + \hat V_0[f_\xi] + \hat V_1[f_\xi]\nabla + \hat V_2[f_\xi] \nabla^2\,
\end{align}
%It should be noted that, for $\xi=0$, it is easy to regularize integrals such that
%states with \emph{outgoing} boundary conditions satisfy \eqref{normgood}.

%It is easy to see that the kinetic term $\hat T = \frac{1}{2} \hat P^2$ is unchanged where $F_\xi(x)=x$.
%In addition, the potential operator has to be evaluated in the scaled coordinate $V \rightarrow V(F_\xi)$.

%The $\theta$-independence of 
%Being an arbitrary remapping of coordinates, 
Since the transformation \eqref{eq:GenCS} has to preserve physical quantities,
%In addition to their scaling behaviour, 
physically meaningful states are expected to %have thus also to
be eigenstates of the transformed Hamiltonian $H_\xi(\hat P_\xi,\hat X_\xi) \equiv T[f_\xi] + V(F_\xi)$, such that
%have energy independent
%to any a
%bound and resonant states with  energies, 
%which holds 
%within the CSM, generalizes to the independence with respect to any arbitrary 
%parameter $\xi$ entering the Hamiltonian $H_\xi \equiv T[f_\xi] + V(F_\xi)$:
\be\label{eq:GenCVT}
0=\frac{\partial }{\partial 
\xi}\frac{\braket{\psi_\xi | \hat H_{\xi}| \psi_\xi}}{\braket{\psi_\xi | \psi_\xi}}   =
\frac{\braket{\psi_\xi | 
\frac{\partial \hat H_{\xi}}{\partial\xi} | \psi_\xi}}{\braket{\psi_\xi | \psi_\xi}} \,,
\ee
where $T=P^2/2$, $\ket{\psi_\xi}$ and $\bra{\psi_\xi}$ are the right and left eigenvectors respectively, 
and the last equality derives from the Hellmann-Feynman theorem. 
%Eq. \eqref{eq:GenCVT} allows us to identify an operator whose expectation value should be
%zero on states whose energy does not depend on the coordinate parametrization.
%Indeed, b
Eq.~\eqref{eq:GenCVT} is of course also valid for physical eigenstates with complex energy.
%, provided that it is possible to 
%define a regularized state $\tilde \psi_\xi = \psi_\xi/\sqrt{\braket{\psi_\xi |\psi_\xi}}$ that is 
%also eigenstate of $H_\xi$. 
By expressing 
\begin{align}
&\braket{\psi_\xi | \frac{\partial \hat H_{\xi}}{\partial\xi} | \psi_\xi} = \\ &\int \dd u \left[ \dot{f_\xi}(u) 
\braket{\psi_\xi | \frac{\delta \hat T[f_{\xi}]}{\delta f_\xi(u)} | \psi_\xi} + \dot{F_\xi}(u) 
\braket{\psi_\xi | V'(F_{\xi}(u)) | \psi_\xi} \right]\;, \notag
\end{align}
we can identify
an operator whose expectation value has to be
zero on states having an energy that is invariant under
%that does not depend on the 
reparametrizations~\eqref{eq:GenCS}:
% the following definition of the Generalized Virial Operator:
\be\label{eq:dH/dxi_M}
\frac{\partial \hat H_\xi}{\partial \xi} = 
\hat U[F_\xi,f_\xi] + \hat U_1[f_\xi] \overrightarrow{\partial_x} + 
\overleftarrow{\partial_x}\hat U_{11}[f_\xi] \overrightarrow{\partial_x} + 
\hat 
U_2[f_\xi] \partial_x^2\,
\ee
where
\ba
U[F,f] &\equiv& V'(F) \dot{F} + \frac{1}{2} \dot{f} 
\left\{\frac{{f'}^2}{f^5}-\frac{1}{2}\frac{f''}{f^4}\right\} \,,
% \\
% U_1[f](x) &\equiv&  \dot{f}(x) \frac{f'(x)}{f(x)^4} -  
% \frac{1}{4}\frac{\dot{f'}(x)}{f(x)^3} \,,
\\
U_1[f] &\equiv& -\frac{1}{2} \frac{\dot{f} f'}{f^4}\;,\; U_{11}[f]\equiv
-\frac{1}{2}\frac{\dot{f}}{f^3} \,,\;
U_2[f] \equiv \frac{1}{2}\frac{\dot{f}}{ f^{3}}\,. \notag
\ea
% \ba
% U[F](x) &\equiv& - \frac{1}{2} V'(F(x)) \dot{F}(x)\,,
% \\
% U_0[f](x) &\equiv&  - \frac{1}{4} \dot{f}(x) 
% \left\{\frac{f'(x)^2}{f(x)^5}-\frac{1}{2}\frac{f''(x)}{f(x)^4}\right\} \,,
% % \\
% % U_1[f](x) &\equiv&  \dot{f}(x) \frac{f'(x)}{f(x)^4} -  
% % \frac{1}{4}\frac{\dot{f'}(x)}{f(x)^3} \,,
% \\
% U_1[f](x) &\equiv& \frac{1}{4} \dot{f}(x) \frac{f'(x)}{f(x)^4}\;, U_{11}[f](x)\equiv
% \frac{1}{4}\frac{1}{f(x)^3} \,,
% \\
% U_2[f](x) &\equiv& \frac{1}{2}\frac{\dot{f}(x)}{ f(x)^{3}}\,,
% \ea

All terms can be evaluated in closed-form, except for $V'(F_\xi)$.
However, in practical applications of the \met, it should not be necessary to evaluate $V'$,
%the derivative of the potential, 
as $F_\xi$ is designed such that $\dot{F_\xi} \neq 0$ only where $V'=0$.

%as $\dot{F_\xi}=0$ where $F_\xi=x$.
The scaling behaviour of wavefunctions suggests that it exists another operator
enabling us to single out physical states.
Since $\hat H_\xi$ stems from
the transformation of $\hat X$ and $\hat P$ in Eq.~\eqref{eq:H_SES(M)}, a physically meaningful state $\ket{\psi_\xi}$ has %therefore
to be compatible with the canonical commutation relation of Quantum Mechanics.
In other terms, for the (regularized) normalization $\braket{\psi_\xi | \psi_\xi}$ to be consistent, the relation
%$ \braket{\psi_\xi | \psi_\xi} = -\ii \bra{\psi_\xi} \left[ \hat X_\xi , \hat P_\xi \right] \ket{\psi_\xi} $
$ \bra{\psi_\xi} \left[ \hat X_\xi , \hat P_\xi \right] \ket{\psi_\xi} = \ii \braket{\psi_\xi | \psi_\xi} $
%=
%\braket{\psi_\xi | \frac{F_\xi}{f_\xi} \overrightarrow{\partial_x} + 
%\overleftarrow{\partial_x} \frac{F_\xi}{f_\xi} -F_\xi \frac{f'_\xi}{f_\xi^2} \psi_\xi}  
should hold within the chosen regularization scheme.
Thus, a \emph{physical} eigenstate must \emph{also} satisfy
\be \label{CCVT1}
\frac{\braket{\psi_\xi | \hat O_{\xi}| \psi_\xi}}{\braket{\psi_\xi | \psi_\xi}} =0\;,\;
\hat O_\xi = 1-F_\xi \frac{f'_\xi}{f_\xi^2} +\frac{F_\xi}{f_\xi} \overrightarrow{\partial_x} +
\overleftarrow{\partial_x} \frac{F_\xi}{f_\xi}\;,
\ee
where $\hat O_\xi \equiv 1 + \ii \left[ \hat X_\xi, \hat P_\xi \right]$.
If $\ket{\psi_\xi}$ has no boundary terms, the above condition is evidently satisfied and no explicit 
regularization is needed.
It should be noted that, even for $\xi=0$, \emph{no regularization} is possible for continuum states % normalizations cannot be regularized such that 
to satisfy 
Eq.~\eqref{CCVT1}.% holds.

Moreover, it can be shown that, for any eigenstate of $H_\xi$, Eq.~\eqref{CCVT1} implies Eq.~\eqref{eq:GenCVT}.
%From a physical point of view, it is interesting to associate this invariance to a conserved quantity.
%It is interesting to apply these consideration to the eigenstates of the unscaled hamiltonian.
For instance, let us consider a dilation of the 
original Hamiltonian $F_\lambda(x) =\left. e^{\lambda} x\right|_{\lambda=0}$.
%\eqref{eq:dH/dxi_M} for the original hamiltonian.
%This can be done with $F_\theta(x) =\left. e^{\ii \theta} x\right|_{\theta=0}$. 
We have
\be \label{CCVT0}
\left.\frac{\partial \hat H_\lambda}{\partial \lambda}\right|_{\lambda=0} \!\!\!\stackrel{\text{Eq.\eqref{eq:dH/dxi_M}
}}{=} \!\!- T -\frac{1}{2} \overleftarrow{\partial_x} \overrightarrow{\partial_x} + x V'
\stackrel{\left[ \hat X , \hat P \right] = \ii}{=} \left[ \hat W , \hat H \right] \;,
\ee
where $\hat W =\frac{1}{2} \left( \hat X \hat P + \hat P \hat X \right)=-\frac{\ii}{2} \left( x \overrightarrow{\partial_x} - \overleftarrow{\partial_x} x\right) 
$ is
the Weyl-quantized form of the dilation generator $\mathbf{x \cdot p}$.
The classical virial theorem shows that the latter quantity is conserved on stationary orbits.
The second member, which has to be zero on physical states, 
corresponds to the operator already presented in Ref.~\cite{CGDD2013}, called for this reason Complex Virial Operator.
In a numerical computation, its expectation value is related to the ``pressure'' exerted by the state on the boundaries of the simulation domain.
%
% As physical states should not depend on computational parameters such as the simulation box size, 
% the degree of fullfillment of Eq.~\eqref{eq:GenCVT} can thus be used as an indication of \emph{quality} of the 
% computational setup.
%
%quantity of Eq.~\eqref{eq:dH/dxi_M} is therefore a generalization of the Complex Virial Theorem
%Indeed Eq.~\eqref{eq:dH/dxi_M} may be though
%For a eigenstate $\psi$ of a real hamiltonian $\langle \psi | W | \psi \rangle=0$.
%
Being a commutator with $\hat H$, the last member of Eq.~\eqref{CCVT0} is of course zero on \emph{any} eigenstate of $\hat H$.
However,  Eq.~\eqref{CCVT0} only holds when evaluated on states satisfying Eq.~\eqref{CCVT1}:
on such states the operator $\hat W$ is a truly conserved quantity.
This condition can be easily generalized to arbitrary transformations:
for any choice of $F_\xi$, each eigenstate of $\hat H_\xi$ satisfying Eq.~\eqref{CCVT1}
%\begin{equation}
%0= \langle \psi | \left[ \hat X , \hat P \right] - \ii |\psi \rangle = -\ii \langle \psi | x \overrightarrow{\partial_x} + \overleftarrow{\partial_x} x + 1 |\psi \rangle\;,
%\end{equation}
will have a $\xi$-independent energy.
%  A \emph{physical} eigenstate 
% of $\hat H_\xi$ must be such that the operator
% \be \label{CCVT1}
% \hat W_\xi = 1-F_\xi \frac{f'_\xi}{f_\xi^2} +\frac{F_\xi}{f_\xi} \overrightarrow{\partial_x} + \overleftarrow{\partial_x} \frac{F_\xi}{f_\xi}
% \ee
% has a zero expectation value. Position and momentum operator respect on this state the usual algebra.

The operators in Eqs.\ (\ref{eq:dH/dxi_M}, \ref{CCVT1}) constitute the main results of this paper.
The quantity $\langle \psi_\xi |\hat O_\xi |\psi_\xi\rangle$ is indirectly associated with the
variation of the energy $\langle \psi_\xi |\hat H_\xi |\psi_\xi\rangle$ induced by the coordinate mapping~\eqref{eq:GenCS}.
%with respect 
%to complex absorbing boundaries induced by the \met.
Eq.~\eqref{CCVT1}  can therefore be used as an alternative to Eq.~\eqref{eq:GenCVT}, with the advantage that no
explicit derivative with respect to any of the $\xi$ parameters is needed.
In both cases, the fulfillment of the equation provides an actual criterion for 
distinguishing, {\it a posteriori}, Siegert states from continuum states. 

As physical results should not depend on computational parameters such as the simulation box size, 
the degree of fullfillment of Eqs.~(\ref{eq:GenCVT}, \ref{CCVT1}) in numerical treatments provides an indication of the \emph{quality} of the 
computational setup.
Matricial representations of these equations can be easily evaluated in \emph{any} basis set; 
we point out, however, that 
compositions of operators have to be done explicitly \emph{before} projection onto the basis set.

\begin{figure}
\includegraphics[width=0.45\textwidth]{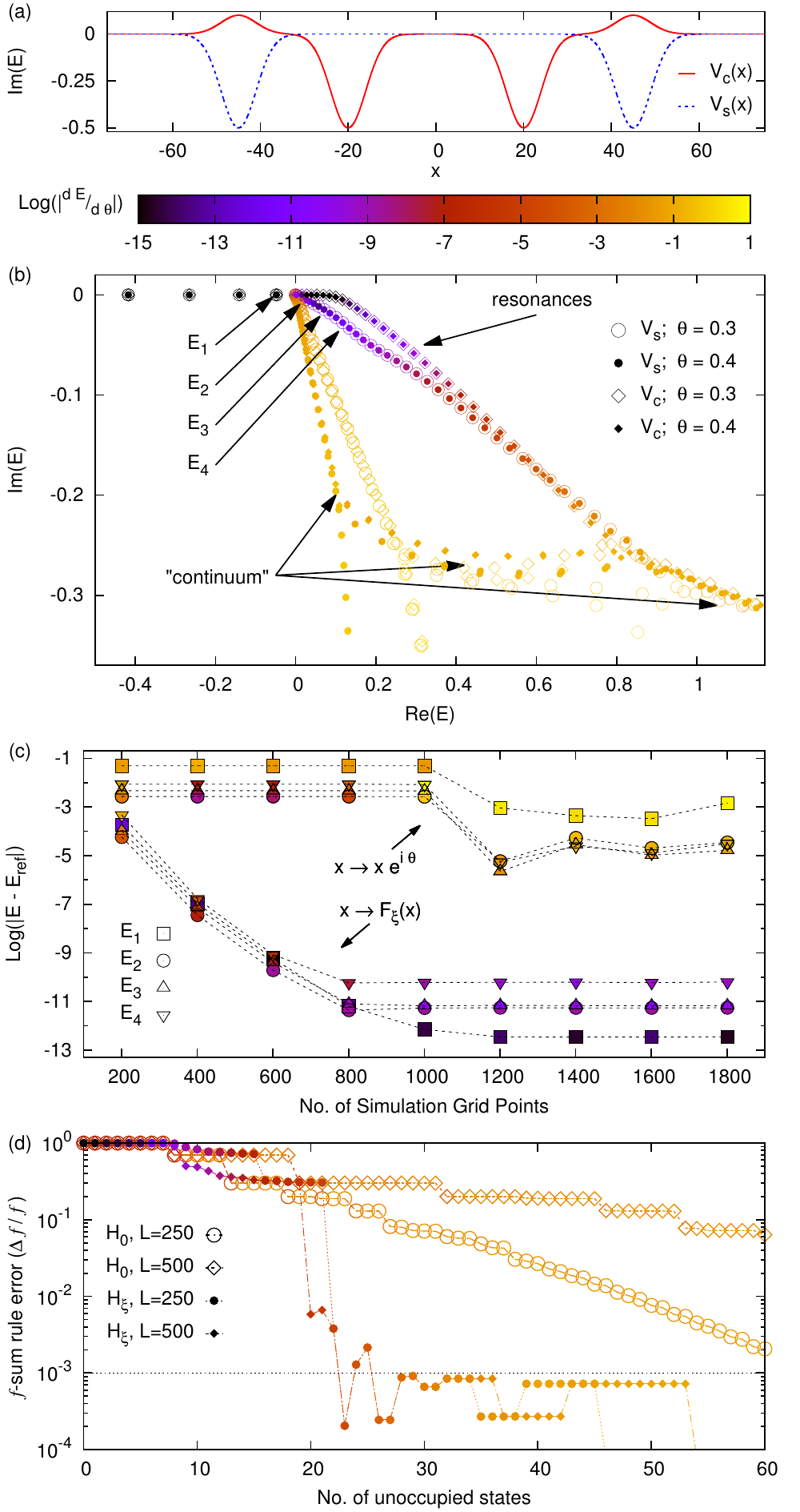}
\medskip
\caption{\label{fig:virial} 
(a) 1D model potentials of Eq.~\eqref{eq:model_potential_full}.
(b) Spectra obtained with different \met\ 
tranformation, compared with the corresponding value of $\frac{\partial E}{\partial \theta}$ from 
Eq.~\eqref{eq:dH/dxi_M}.
The simulation box considered has size $L=300$ AU.
(c) \met\ is compared against CSM for the Hamiltonian with $V_s$ potential. 
The error in the identification of some resonant energies (pointed by the arrows in panel (b)) is plotted against 
the number $N$ of degrees of freedom used in the simulation. Reference values are extracted with the \met\ at $N=3000$.
(d) Fulfillment of the $f$-sum rule for the $V_c$ potential
as a function of the number of empty states considered for different $L$,
ordered by $\left|\frac{\partial E}{\partial \theta}\right|$.
Fermi level is chosen such as to have 8 occupied states.
}
\end{figure}

% The evaluation of the above quantities allows therefore one to identify the important 
% states that can be extracted from a computational treatment of non-hermitian hamiltonians originated
% from \met.
% When the computational treatment is accurate enough, 
% resonant and bound states can be identified with lowest-valued of the above quantity.
In order to demonstrate the significance of the information obtained from Eq.~\eqref{eq:dH/dxi_M},
%together with a comparison of the \met\ and the CSM, 
we first discuss its numerical application to one-dimensional potentials, made up by superpositions of Gaussian functions as
indicated in \eqref{gaussian}. Being $v(x_c,x) = -\frac{1}{2}\left[G(4,-x_c;x) + G(4,x_c;x)\right]$, we define (see 
Fig.~\ref{fig:virial}-a)
%described by the following  potential:
\be\label{eq:model_potential_full}
V_s(x) = v(45,x)\;, \quad V_c(x) = v(20,x) - 0.2 V_s(x)\,.
\ee
The Hamiltonian with $V_s$ models two diffusive centers far from each other, yet close enough to exhibit a non-trivial scattering structure.
%On the other hand 
The other toy potential $V_c$, tailored to
preserve the degeneracy of the bound eigenspaces, models a charge transfer from a ``core'' region,  to the
``shell'' potential $V_s$.
These two systems therefore present very similar bound states
but rather different spectra of resonant states.
The one-dimensional Hamiltonians are discretized in a Daubechies wavelet basis
set as described in Ref.~\cite{CGDD2013}. The \met\ method is applied~\setcounter{footnote}{99}\footnote{We use here
$F(x)= x \exp(\ii \theta g(x)/2)$, where $g(x)=1+  \mathrm{erf}(\lambda (x-x_0)) -\mathrm{erf}(\lambda (x+x_0))$, $\lambda=0.12$ and $x_0=95$.}.

Fig.~\ref{fig:virial}-b provides the evidence that bound and 
resonant states are stable with respect to variations
of the scaling parameter $\theta$. This fact is clearly confirmed by the corresponding value of 
$\frac{\partial E}{\partial \theta}$, obtained from Eq.~\eqref{eq:dH/dxi_M}. 
In Fig.~\ref{fig:virial}-c, we show that for the potential $V_s$  the \met\ outperforms the CSM.
Since the Gaussians are far from each other, in case of the CSM the potential
and the eigenfunctions oscillate so strongly that
very small grid spacings are required for the reliable identification of resonant energies.
 %When the grid spacing is not small enough, the Complex Scaled potential is 
%so inaccurate that the resonant energies are \emph{badly} positioned, leading to wrong results. 
%Moreover 
%Even when the potential is correctly discretized, the interesting resonances
%might be 
%eigenstates of the hamiltonian in the desired energy range might oscillate even more than the potential.
%The number of degrees of freedom has to be considerably increased to reveal such a situation.
Even with grids five times denser than the one used at $\theta=0$ -- already dense enough to correctly 
discretize $V_s(xe^{\ii \theta})$ --  %, see Fig.~\ref{fig:kstar_vs_theta}), 
the CSM still provides incorrect estimates of resonant energies.
This fact is further confirmed by the corresponding values of the complex virials.
%of $\theta$, $x_0$, $\lambda$. In order to do so, we have introduced the 
%following averaged quantity:
%\be
%\Delta E \equiv \frac{1}{3}\sum_{\xi = \theta, x_0, \lambda}\left| 
%\frac{\partial E}{\partial \xi} \right|\,.
%\ee
\begin{figure}[h!]
\includegraphics[width=0.5\textwidth]{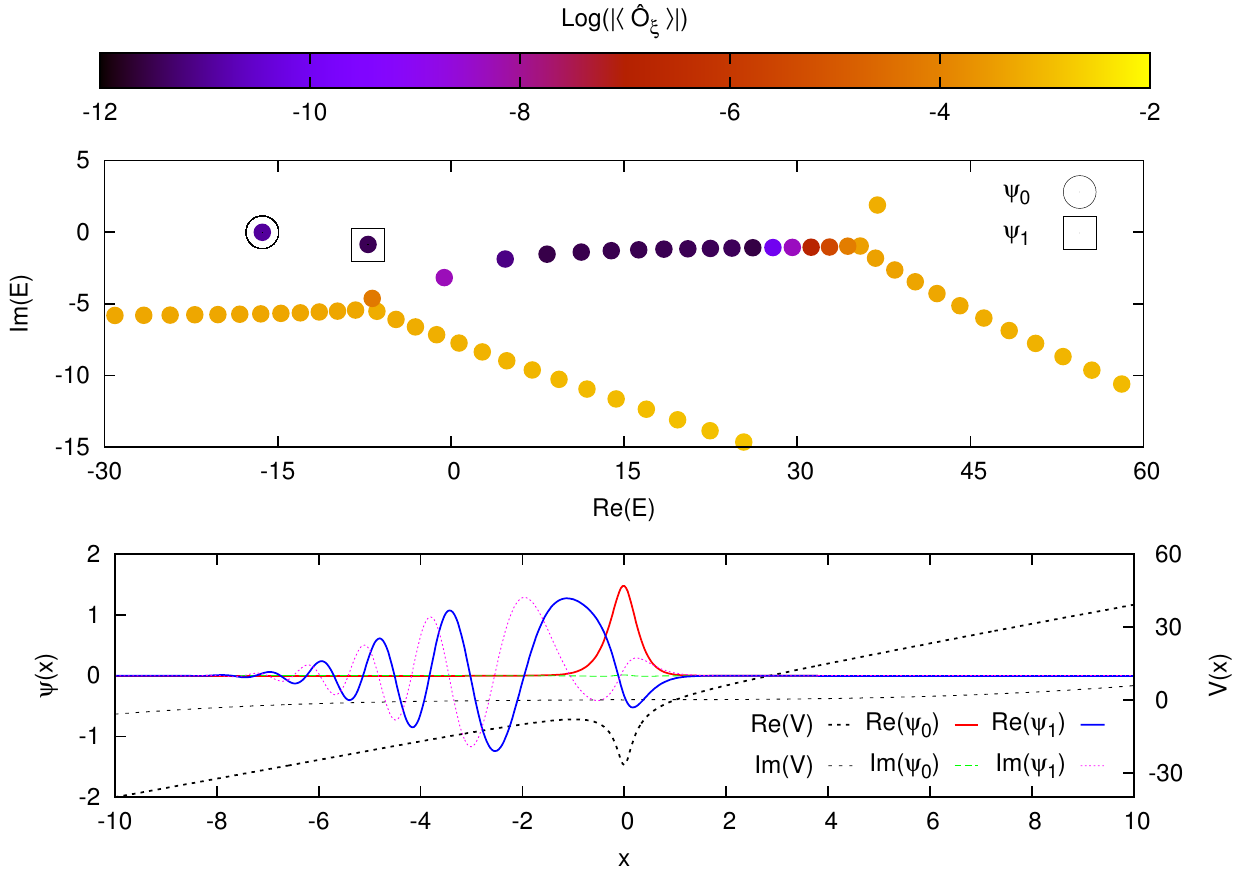}
%\medskip
\caption{\label{fig:CVT} Measurement of $\langle \hat O_\xi\rangle$ from Eq.~\eqref{CCVT1} for each state in the spectrum of a 1D ``soft-Coulomb'' potential with an external electric field.
The parameters used are as in Fig.\ 2 of Ref.\ \cite{doi:10.1021/jz401110h}. 
We used the \met\ \cite{Note100}, with $\theta=0.4$ and $x_0=12$, lying therefore 
\emph{outside} the simulation box ($L=10$ AU).
This enables us to extract the \emph{unscaled} 
form of the true resonant eigenfunctions corresponding to the perturbation of the lowest bound states, indicated by 
$\psi_0$ and $\psi_1$.}
\end{figure}

As mentioned in the introduction, resonant states can be used to obtain a compact 
representation of the Green's function. 
In systems with open boundaries, such a compact representation is very useful to express
linear response functions and the derived optical spectroscopic properties in an optimal way.
A good indicator of the quality of a discrete basis set for optical properties is the fulfillment of the 
Thomas-Reiche-Kuhn sum rule (or $f$-sum rule, see e.g. Ref.~\cite{tddft2}), which relates the first momentum 
of the oscillator strengths to that of the equilibrium density of the system, 
i.e.\ the number of states below the Fermi level.
The information provided by Eq.~\eqref{eq:dH/dxi_M} is of great utility in this case:
%the states with lowest value of the virial, 
%being closer to true Siegert states, encode the correct analytic structure of the resolvent, responsible of
%the linear response of the system. 
%To give a quantitative evidence of this point, 
we have plotted in Fig.~\ref{fig:virial} the fulfillment of the $f$-sum rule as a function of 
the number of unoccupied states considered, ordered by
$\frac{\partial E}{\partial \theta}$, for
%the
%states with lowest virial, of 
the Hamiltonian with the $V_c$ potential. 
With such ordering, we are able to identify the \emph{minimal} set of states allowing the 
fulfillment of the sum rule up to an excellent accuracy, \emph{independently} of the simulation box 
size. This cannot be achieved with the low-energy eigenstates of the original, unscaled Hamiltonian,
which suffer from the well-known continuum collapse~\cite{Natarajan201229}.
Equivalent results can also be obtained by using the operator of Eq.~\eqref{CCVT1}.

%Such a criterion 
Our approach can also be used to identify field-induced metastable states of Hamiltonians in non-trivial environments.
A notable example of such metastability is given by the Stark ionization of molecules,
recently studied within the framework of Density Functional Resonance Theory~\cite{Whitenack2011}.
The reliable determination of metastable states that have to be occupied is admittedly a problem in the latter treatment.
However, our method provides a natural solution to this problem.
%easily generalizable beyond the uniform complex scaling.
As an example, in Fig.~\ref{fig:CVT}, we show the values of $\langle \hat O_\xi \rangle$ of Eq.\ \eqref{CCVT1} for a model 
Hamiltonian with a ``soft-Coulomb'' 1D potential under external electric field of intensity $\mathcal E$, see e.g. Ref.~\cite{doi:10.1021/jz401110h}.
States with smallest $|\langle \hat O_\xi \rangle|$ are those originating from a perturbation of the
Rydberg states at $\mathcal E=0$,
%The ranking of eigenstates provided by the GCVT is of great utility in considering the metastable states
therefore not belonging to the set of ``continuum'' states of the \met\ Hamiltonian.
%As a perspective, t
This approach would be very helpful in identifying the \emph{physical} metastable 
channels of open quantum devices under external electric fields~\cite{SourceSink}.

To summarize, we have presented a simple and robust method to numerically identify the Siegert states of  open 
quantum systems undergoing a \emph{generic} coordinate reparametrization.
Such method, based on rigorous analytic derivations, makes the usage of rescaled Hamiltonians
much more powerful than the artificial addition of {\it ad hoc} Complex Absorbing Potentials to the device.
This method is especially useful for coordinate mappings leaving the potential of the quantum device unscaled,
and does not require any finite-difference measurement of the eigenvalue sensitivity, avoiding the need of 
tracing trajectories in the parameter space and to search, graphically, for stable points.
Our findings, supported by numerical examples with 1D model Hamiltonians, % but
can straightforwardly be extended to 3D systems using  e.g.\ separable forms of coordinate 
mappings~\cite{:/content/aip/journal/jcp/127/3/10.1063/1.2753485},
%Being based on analytic derivations, our method is therefore completely general and 
and can be applied to \emph{any} numerical basis set.
The method provides \emph{scalar quantities}
%natural ``ranking'' of states with complex energies, 
enabling the identification of %indicating which 
%useful in determine which 
%states are 
physically relevant states and, at the same time, indicates the reliability of the computational setup.
%Thanks to this method, we believe that hamiltonian with complex absorbing boundaries might be 
%more easily treated, 
We believe that our method paves the way for accurate investigations of scattering properties of open quantum systems. 
Work is in progress towards the iterative extraction of Siegert states out of non-Hermitian Hamiltonians.

The authors thank I.\ Duchemin for useful discussions.
%
%\bibliography{RSthroughGenCS}
%\include{biblio}
%

\end{document}